\documentclass[preprint2,twocolumn,times,tighten]{aastex63}

\usepackage{graphicx,times}
\usepackage{subfigure}
\newcommand{\be}{\begin{equation}}
\usepackage{threeparttable}
\usepackage{booktabs}
\newcommand{\ee}{\end{equation}}
\newcommand{\bea}{\begin{eqnarray}}
\newcommand{\eea}{\end{eqnarray}}
\newcommand*{\sy}{\color{black}}

\newcommand\cmtch[1]{{\color{red}[CH: #1]}}

\usepackage{bigints}
\usepackage{amsmath}
\usepackage{cases}
\usepackage{longtable}
\usepackage{hyperref}
\usepackage{epstopdf}
\usepackage{amsmath,bm}
\usepackage{amssymb}
\usepackage{natbib}
\usepackage{morefloats}
\usepackage{multirow}
\usepackage{array}
\usepackage{verbatim}
\usepackage{color}
\usepackage{lineno}
\newcommand{\red}[1]{\textcolor{black}{#1}}
\newcommand{\blue}[1]{\textcolor{blue}{#1}}

\begin{document}

\title{Wide binary stars formed in the turbulent interstellar medium
}

\email{sxu@ias.edu; hchwang@ias.edu; chamilton@ias.edu; dong@astro.cornell.edu}

\author[0000-0002-0458-7828]{Siyao Xu\footnote{Hubble Fellow}}

\affiliation{Institute for Advanced Study, 1 Einstein Drive, Princeton, NJ 08540, USA}

\author[0000-0003-4250-4437]{Hsiang-Chih Hwang}
\affiliation{Institute for Advanced Study, 1 Einstein Drive, Princeton, NJ 08540, USA}

\author[0000-0002-5861-5687]{Chris Hamilton}
\affiliation{Institute for Advanced Study, 1 Einstein Drive, Princeton, NJ 08540, USA}

\author[0000-0002-1934-6250]{Dong Lai}
\affiliation{Center for Astrophysics and Planetary Science, Department of Astronomy, Cornell University, Ithaca, NY 14853, USA}

\begin{abstract}

The ubiquitous interstellar turbulence regulates star formation and 
the scaling relations between the initial velocity differences and the initial separations of stars. {We propose that} 
the formation of wide binaries with initial  separations {$r$} in the range 
%$0.01~\text{pc} <r<1$ pc (i.e.,
$\sim 10^3~\text{AU} \lesssim r \lesssim 10^5$~AU is a 
natural consequence of star formation in the turbulent interstellar medium. 
With the decrease of $r$, 
the mean turbulent {relative} velocity $v_\text{tur}$ between a pair of stars decreases, while the largest velocity $v_\text{bon}$ {at which they still may} be gravitationally bound increases. 
\red{When $v_\text{tur} < v_\text{bon}$, a wide binary can form.} 
In this formation scenario, 
%we find that both the probability for wide binary formation and the wide binary fraction in a star-forming region strongly depend on the turbulence properties. 
we derive the eccentricity distribution $p(e)$ of wide binaries for an arbitrary \red{relative velocity distribution.
By adopting a turbulent velocity distribution},
we find that wide binaries at a given initial separation generally exhibit a superthermal $p(e)$. {This provides a natural explanation for the} observed superthermal $p(e)$ of the wide binaries in the Solar neighbourhood. 
%Turbulence also affects the wide binary fraction in a star-forming region, which can be highly inhomogeneous, depending on the local turbulence properties. 

\end{abstract}

%\keywords{}

\section{Introduction}

Gaia 
\citep{Gaiadr1,Gaiadr2}
has revealed a large number of wide binaries with semimajor axes $a \gtrsim 10^3~$AU
(e.g., \citealt{Elk18,Igo19,Hwang20,Tian20,Hart20,Elb2021}).
Due to their sensitivity to gravitational perturbations, 
wide binaries have been used to probe unseen 
{Galactic disk} material
\citep{Bah85},
dark matter substructure 
\citep{Pena16},
MAssive Compact Halo Objects (MACHOs,
\citealt{Chaname2004,Quinn2009,Monr14}),
and constrain the dynamical history of the Galaxy
\citep{Alle07,Hwang21b}. 
%The {presence of a wide binary companion} can also significantly influence {evolution of planetary systems}
%\citep{Kaib13}.

Despite their important astrophysical implications, the origin of wide binaries remains a mystery. 
Due to their large $a$'s (comparable to the typical size $\sim 0.1~$ pc 
$\approx 2\times10^4$ AU
of molecular cores, \citealt{Ward07})
and the dynamical disruption in star clusters, 
it is believed that very wide binaries with $a\gtrsim 0.1~$pc can hardly form and survive in dense star-forming regions
\citep{Deac20}. 
Various wide-binary formation mechanisms have been proposed, including 
%fragmentation of turbulent molecular cores
%\citep{Good07},
cluster dissolution
\citep{Kouw10, Moeckel2011},
dynamical unfolding of compact triple systems
\citep{Reip12}, formation 
in adjacent cores with a small relative velocity 
\citep{Toko17},
and in the tidal tails of stellar clusters
\citep{Pen21,Liver23}.
Observations of the large fraction of wide pairs of young stars in low-density star-forming regions 
\citep{Toko17},
%the broad period distribution of binaries 
%\citep{Good07}
the chemical homogeneity between the components of wide binaries
\citep{Hawk20},
the metallicity dependence of the wide-binary fraction
\citep{Hwang21},
and N-body simulations 
(e.g., \citealt{Krou01})
\red{suggest that 
%wide binaries are likely 
%to form in a star-forming region and 
wide binary-components are likely to form in the same star-forming region
\citep{Andre19}.} 
%\blue{(HCH: are you trying to say: most wide binaries are formed from the same star formation site (`co-natal'), instead of forming from the pairing in the later stage?)}

%to be an intrinsic property related to star formation,
%rather than a consequence of long-term dynamical evolution or 
%random entrapment of stars in the field
%\citep{Mak12}. \blue{(HCH: need some explanations why these theoretical and observational results have something to do with the concluding sentence?)}

In addition, {\it Gaia}’s
high-precision astrometric measurements reveal 
superthermal eccentricity distribution for wide binaries with separations $\gtrsim 10^3$ AU
\citep{Toko20,Hwang21t}.
\citet{Hamil22} 
{suggests that a superthermal eccentricity distribution cannot be produced dynamically, implying that the initial distribution must itself
be superthermal}.

%(see also, e.g., \citealt{Sada17}). 
%"the further interaction with passing stars and Galactic tide makes the eccentricity distribution closer to the thermal eccentricity distribution"

%The initial binary population [i.e., the population after small-N decay and internal energy redistribution processes, “eigenevolution” (see below), has modified it] is further changed by dynamical interactions within a cluster on a timescale of 1 m.y. for typical galactic star-forming clus- ters (Kroupa, 1995a,b; Adams and Myers, 2001). Encoun- ters will destroy soft and active binaries, leaving mostly the hard binary population unchanged.
%"Almost all young stars are found in multiple systems with a very wide separation distribution"

The interstellar medium (ISM) and star-forming regions are turbulent, 
%\blue{(HCH: should we briefly mention the origin of the turbulence here, just like the beginning of Sec 3? Is turbulence all coming from supernovae? I asked because in binary formation theory, people talk a lot about other turbulence, like turbulent core fragmentation)}, 
with the turbulent energy mainly coming from supernova explosions
\citep{Pad16}.
The observationally measured power-law spectrum of turbulence spans many orders of magnitude in length scales, from $\sim 100$ pc 
down to $\sim 10^{-11}$ pc in the warm ionized medium %\blue{(HCH: Thanks! But I suggest that we use the same length unit throughout the paper. Maybe pc or AU alone?)}
(e.g., \citealt{Armstrong95,CheL10,XuZ17,Xup20})
%\blue{(HCH: down to what scale in pc? I saw some in Sec. 3 and those scales are very useful for intro. That would give the reader some physical scale compared to wide binaries)}
and down to $\sim 0.01$ pc in molecular clouds
(MCs)
\citep{Laz09rev,Hen12,Yuen22}. 
The ubiquitous turbulence 
plays a fundamental role in the modern star formation paradigm 
\citep{MacL04,ElmegreenScalo,Mckee_Ostriker2007}.
Our understanding of star formation has significantly improved thanks to the recent
development in theories of magnetohydrodynamic (MHD) turbulence 
\citep{GS95,LV99,CL02_PRL}, 
MHD simulations 
(e.g., \citealt{Stone98,Fede11,Pad16,Kri17}),
and techniques {for} measuring interstellar turbulence 
(e.g., \citealt{LP00,HB04,LP06,Laz18,Xuy21,Burk21}).

Turbulence not only regulates the dynamics of MCs and the formation of density structures, e.g., 
filaments, clumps, and cores, where star formation takes place, 
but also affects the 
kinematics of molecular gas and dust
\citep{Hen12},
density structures, and young stars over a broad range of length scales. 
Dense cores in MCs 
\citep{Qia18,XuS20}
and young stars 
\citep{Ha21,Krol21,Zhou21,Ha22}
inherit their velocities from the surrounding turbulent gas, 
and thus the statistical properties of their velocities are similar to that of the turbulent gas.

%If a pair of young stars are gravitationally bound, 
%their initial turbulent velocities and the related kinetic energy are important for determining 
%how widely separated the binary stars can be. 

In this letter,  
%\cmtch{I'd delete from here...} based on the theoretical understanding on the role of turbulence in star formation and 
%motivated by the recent observations confirming the turbulent velocities of young stars \cmtch{... to here.}, 
we will investigate how the initial turbulent velocities between pairs of stars affect the formation of wide binaries and their eccentricities. 
We first describe the initial turbulent velocities of pairs of stars in Section 2. 
In Section 3, we focus on the formation of wide binaries and their eccentricity distribution. 
More discussion is provided in Section 4. 
Our conclusions follow in Section 5.

\section{Initial turbulent velocities of stars}
\label{sec:tur}

%The turbulent energy is injected by supernova explosions on large scales ($\sim 100$ pc,
%\citealt{Cham20})
%and cascades down toward smaller and smaller scales.
%The supernova-driven turbulence provides the source 
%and sustains the level of turbulence observed 
%over the MC lifetime ($\sim 10$ Myr),
%and the MC turbulence controls the cloud formation 
%and dispersion
%\citep{Pad16}. 
%By using cold gas and dust tracers, turbulent velocity and density spectra have been measured in MCs 
%from $\sim 10$ pc to $\sim 0.01$ pc
%\citep{Laz09rev,Hen12,Yuen22}.

Stars form in high-density structures, i.e., filaments, cores, in turbulent MCs, 
which are generated by the 
compressions and shocks in highly supersonic turbulence
\citep{Fed09,Moc18,Ino18,XJL19}.
Dense cores arise at the collision interfaces of converging turbulent flows,
with turbulent motions existing on sub-core scales
\citep{Volg04}.
Naturally, the turbulent velocities of gas are imprinted in those of newly formed stars. 
As confirmed by recent {\it Gaia} observations
(e.g., \citealt{Ha21,Zhou21,Ha22}), 
the velocity differences and spatial separations of young stars 
statistically 
follow the power-law velocity scaling of interstellar turbulence.
%except in high stellar density regions, e.g., the Orion Nebula Cluster, 
%where the initial turbulence scaling relation is rapidly dynamically disrupted.

%\blue{HCH: do we need to worry about any timescale issues? Are these turbulence profile time-dependent (e.g. after supernova explosion? Or cascade time?)?}

%\blue{HCH: Maybe we can start with some intro on our scenario. Stars are randomly distributed (without other binary formations), and every star has an initial velocity following the turbulence $v_t$. Then we investigate how many of them form bound binaries. (Originally I didn't understand what initial velocities and initial separations are for. Adding an intro may be clearer) } 
We consider that in a star-forming region, %the newly formed stars are randomly distributed, and 
the initial velocity differences and spatial separations of stars at birth statistically satisfy the {\sy averaged} turbulent velocity scaling, 
\begin{comment}
%----------------------------------
We assume \blue{(this turns out to be a critical assumption. Can we cite a reference here, so that we don't need to say "We assume"?)} that the initial velocity difference $v_{12,0}$ ($v_{12,0}>0$) between a pair of stars has a 
normalized probability density function (PDF) as 
\begin{equation}\label{eq: norpdf}
    f (v_{12,0}) = \frac{f^\prime(v_{12,0})}{\int_0^\infty f^\prime(v_{12,0}) dv_{12,0}}, 
 %   = \frac{f^\prime(v_{12,0})}{1- F^\prime_{v_{12,0}}(0)}.
\end{equation}
and $f^\prime(v_{12,0})$ follows a Gaussian distribution, 
\begin{equation}
     f^\prime(v_{12,0}) = \frac{1}{\sigma \sqrt{2\pi}} \exp{\bigg[-\frac{1}{2} \Big(\frac{v_{12,0}-\mu}{\sigma}\Big)^2\bigg]}.
\end{equation}
The mean $\mu$ and the standard deviation $\sigma$ of the distribution are both given by the turbulent velocity $v_\text{tur}$,
%----------------------------------
\end{comment}
\begin{equation}\label{eq: genturind}
   v_\text{tur} = \langle v \rangle= V_L \Big(\frac{r}{L}\Big)^\alpha,
\end{equation}
%where $\bm{v}_{t,1}$ and $\bm{v}_{t,2}$ are the initial velocities of the two stars in the observer frame, 
where $r$ is the initial separation between a pair of stars, {\sy and $v$ is the absolute value of their initial velocity difference.} 
%\blue{(need a bit more explanation: Eq. 12 is seen from observations (right?) and therefore $\mu=v_{\rm tur}$ by definition. But we still need to explain why $\sigma=v_{tur}$)}. 
The injected turbulent speed 
$V_L$ and the injection scale $L$ have typical values as 
$\sim 10~ \text{km}~\text{s}^{-1}$ 
and $\sim 100~\text{pc}$ for interstellar turbulence 
\citep{Cham20}.
The power-law index $\alpha$ reflects the properties of turbulence,
which is typically 
$1/3$ for solenoidal turbulent motions with the Kolmogorov scaling, 
and close to $1/2$ for highly compressive turbulent motions dominated by shocks
\citep{Fed09,KowL10}.
A steeper turbulent velocity scaling 
(i.e., a larger $\alpha$)
corresponds to a more efficient energy dissipation in shock-dominated turbulence.
%and a faster decline of $v_\text{tur}$ with decreasing $r$.
%\blue{(HCH: I like this. Does steeper mean a larger alpha?)}
The Kolmogorov scaling applies to {most of the volume of} an MC,
while the steeper scaling is preferentially seen in small-scale high-density regions that result from shock compressions 
\citep{Laz09rev,XuS20,Xuy21,Rani22,Yuen22}.

\begin{figure*}[ht]
\centering
\subfigure[]{
   \includegraphics[width=8.7cm]{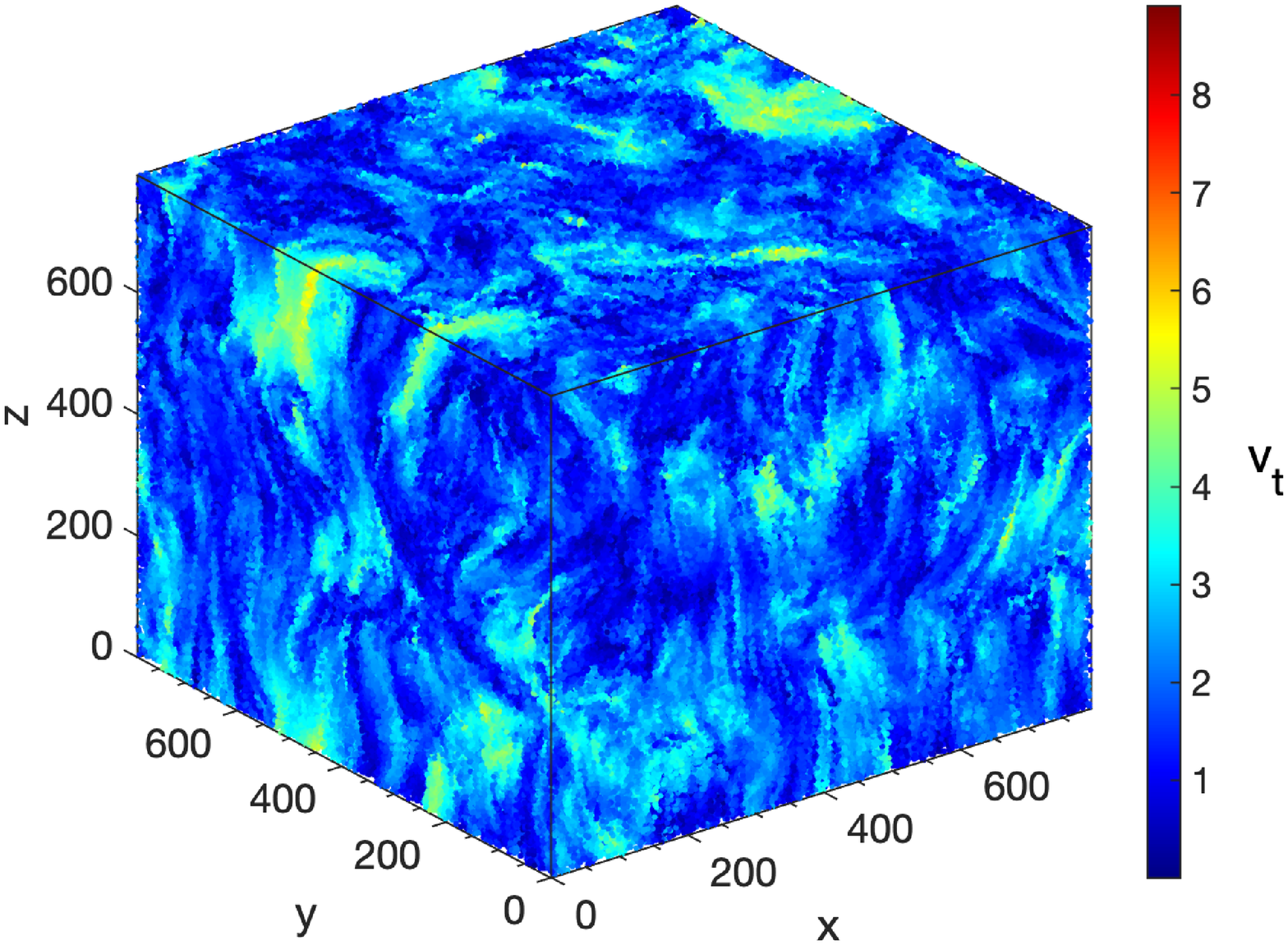}\label{fig: numvmap}}
\subfigure[]{
   \includegraphics[width=8.7cm]{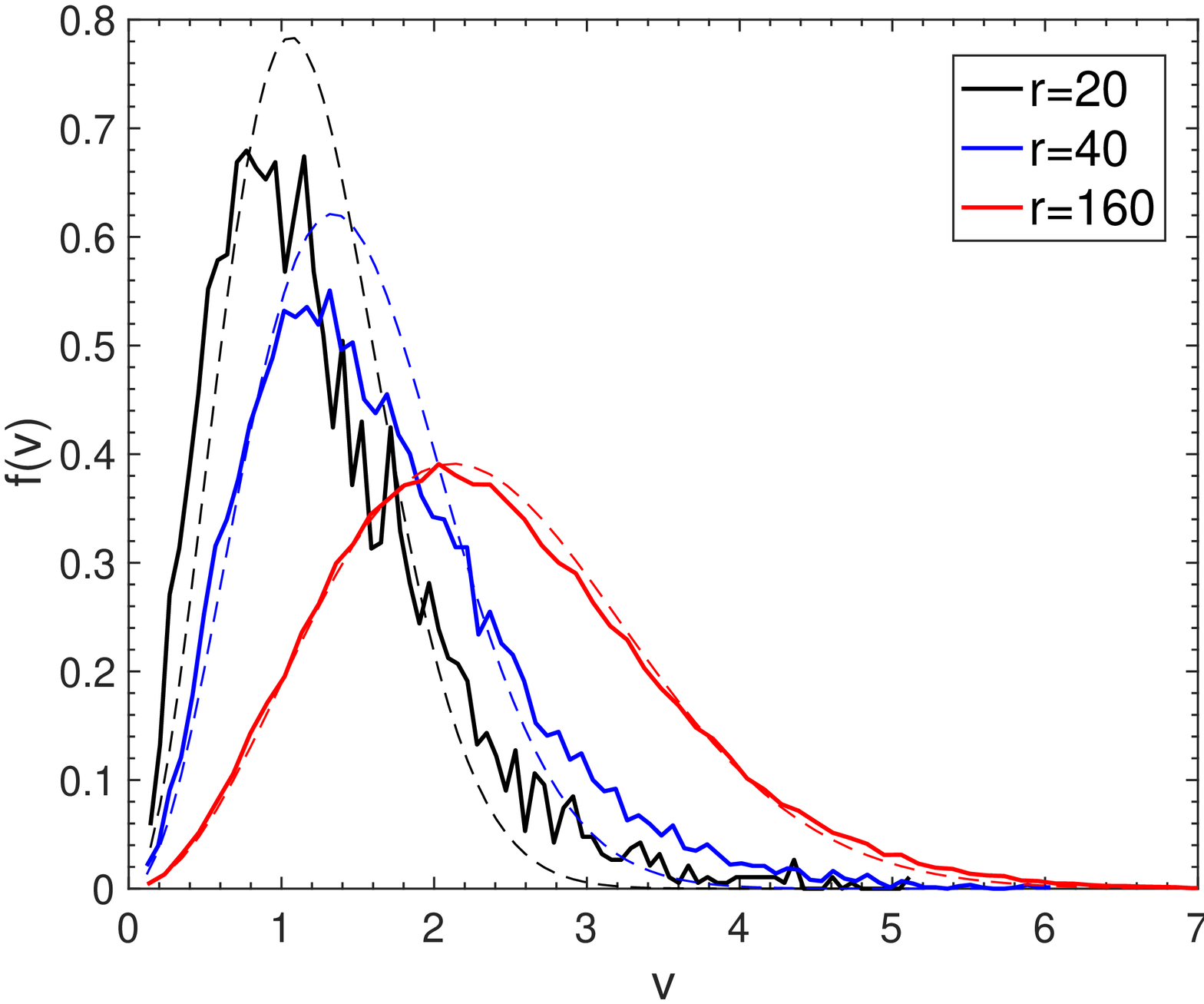}\label{fig: numfv}}
\caption{(a) 3D distribution of gas velocities $v_t$ (in numerical units) taken from an MHD turbulence simulation
\citep{Hux21}.
(b) Turbulent speed distribution measured at different $r$'s (in the units of grids) in the MHD turbulence simulation. The dashed lines correspond to Eq. \eqref{eq: 3dveldis}.}
\label{fig:numsim}
\end{figure*}

\red{In Fig. \ref{fig: numvmap}, we illustrate the 3D distribution of gas velocities $v_t$
\footnote{Note that $v_t$ is the velocity at each point, while $v$ is the velocity difference between two points separated by $r$.}
taken from 
MHD turbulence simulations in physical conditions similar to those of a star-forming region
\citep{Hux21}, with
the sonic Mach number $M_s = V_L /c_s \approx 10$ and Alfv\'{e}n Mach number $M_A = V_L / V_A \approx 0.5$, where $c_s$ is the sound speed, and $V_A$ is the Alfv\'{e}n speed.
The grid resolution is $792^3$,
$L$ is about half of the box size, and $V_L \approx 3 $ (numerical unit). 
By using the turbulent velocities from the simulation,  
we measure the speed distribution $f(v)$ 
(such that $\int f(v) dv =1$)
corresponding to different $r$'s and present it in Fig. \ref{fig: numfv}.
It approximately follows
\begin{equation}\label{eq: 3dveldis}
f(v) \approx 
\sqrt{\frac{2}{\pi}}\frac{1}{\sigma_v^3}\exp{\Big(-\frac{v^2}{2\sigma_v^2}\Big)} v^2.
\end{equation}
We find that 
the parameter $\sigma_v$ 
is comparable to
$v_\text{tur}$ (Eq. \eqref{eq: genturind}). %\cmtch{Can we justify that statement briefly, add a reference?} \blue{(HCH: should use speed distribution?)}
The dashed lines in Fig. \ref{fig: numfv} represent $f(v)$ given by Eq. \eqref{eq: 3dveldis} with
$\sigma_v \approx 0.7 v_\text{tur} $ and $\alpha = 1/3$, which approximately agree with the numerical results.} %\blue{HCH: so it's Maxwellian distribution, right? I wouldn't call $\sigma_v$ as "standard deviation", because the standard deviation of $f(v)$ is $\sigma_v\sqrt{(3\pi-8)/\pi}$ (from wiki), not $\sigma_v$. Maybe just call it `parameter', or `width parameter'?}

We note that due to different effects in star-forming regions, 
e.g., protostellar outflows,
turbulent velocities may not always follow a power-law scaling
\citep{HuF22}.
In addition, 
we will adopt a single turbulent scaling only down to $\sim 0.01~$ pc ($\sim 2000$ AU)
as it may not apply on smaller scales due to the gravitational compression and gravity-driven turbulence
\citep{Xug20,Guer20}.
We also ignore the fact that {turbulence in the presence of magnietic fields is typically anisotropic} 
\citep{GS95,LV99}. {The} impact {of these additional effects upon} binary formation will be investigated in our future work.

\section{Wide binaries with initial turbulent velocities}
\label{sec:fortur}

\subsection{Formation of wide binaries}

\begin{figure}[ht]
\centering
%\subfigure[$M=2 M_\odot$]{
   \includegraphics[width=9cm]{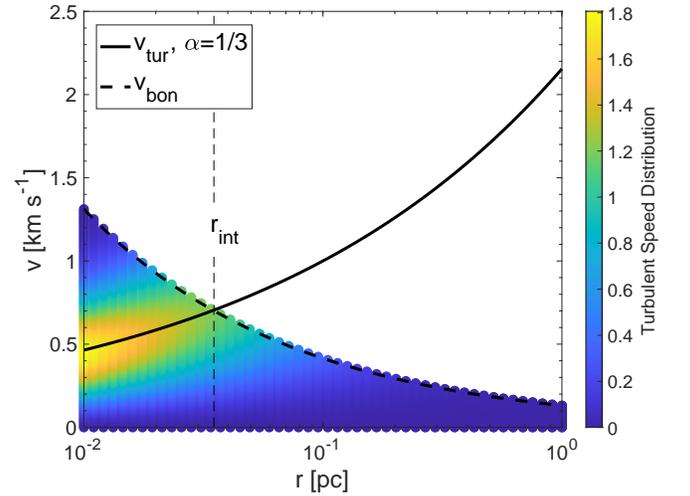}
%\subfigure[$M=10 M_\odot$]{
%   \includegraphics[width=8.7cm]{vr12M10a3.eps}\label{fig: m10a3}}
%\subfigure[$M=2 M_\odot$]{
%   \includegraphics[width=8.7cm]{vr12M2a2.eps}\label{fig: m2a2}}
%\subfigure[$M=10 M_\odot$]{
%   \includegraphics[width=8.7cm]{vr12M10a2.eps}\label{fig: m10a2}}
\caption{Initial velocity difference $v$ and separation $r$ between two stars formed in a turbulent environment. 
The colored region corresponds to 
a gravitationally bound system 
with $M=2 M_\odot$.
The color scale indicates the turbulent
speed distribution {(Eq. \eqref{eq: 3dveldis})}.
The thick dashed and solid lines 
correspond to $v_\text{bon}$ (Eq. \eqref{eq: gengbcon}) and $v_\text{tur}$ (Eq. \eqref{eq: genturind}), respectively. %\blue{(I may just keep one panel, and use some texts to explain the expected changes for other parameters?)} 
The vertical dashed line corresponds to $r_\text{int}$ (Eq. \eqref{eq:rint}).
The parameters $V_L= 10~ \text{km}~\text{s}^{-1}$, $L= 100~\text{pc}$, and $\alpha = 1/3$ are used. }
\label{fig: vr12}
\end{figure}

For a pair of stars to be gravitationally bound at birth, the relative velocity
$v$ and $r$ must satisfy 
%(Eq. \eqref{eq: covegerm}, see Appendix \ref{sec:app})
\begin{equation}\label{eq: gengbcon}
    v < {v_\text{bon}(r) \equiv}  \sqrt{\frac{2GM}{r}},
\end{equation}
where 
%we define $v_\text{bon}$ as the binding velocity of a gravitationally-bound binary system, 
%$G$ is the gravitational constant,
%and 
$M$ is the total mass of the binary. 
%\blue{(maybe remove Eq. 13 and just refer to Eq 9?)}. 
As an illustration, 
in Fig. \ref{fig: vr12}
we present $v$ vs. $r$ for pairs of stars formed in turbulent gas 
over the range $0.01~\text{pc} <r<1$ pc (i.e., $2\times10^3~\text{AU} < r< 2\times10^5$~AU).
%The color scale indicates the 3D velocity distribution, 
%{which we \red{approximate as} a Gaussian 
%\citep{Towns47,Wilc11},
% fig 4.1
%in comparison with the relation in Eq. \eqref{eq: gengbcon}.
%that follow the above condition for different values of $M$. 
The colored region corresponds to 
$v<v_\text{bon}$ for $M=2 M_\odot$.
\red{The color scale corresponds to $f(v)$ given by Eq. \eqref{eq: 3dveldis}.}

The intersection between $v_\text{bon}$ and $v_\text{tur}$ occurs at 
\begin{equation}\label{eq:rint}
    r_\text{int} 
    = \Big(\frac{2 GM L^{2\alpha}}{V_L^2}\Big)^\frac{1}{2\alpha+1},
\end{equation}
which increases with increasing $M$ and $\alpha$.
For instance, at $\alpha=1/3$ we have 
\begin{equation}\label{eq:rintk}
    r_\text{int} \approx 0.023~\text{pc} 
    \Big(\frac{M}{M_\odot}\Big)^{0.6} \Big(\frac{L}{100~\text{pc}}\Big)^{0.4} \Big(\frac{V_L}{10~\text{km s}^{-1} }\Big)^{-1.2},
\end{equation}
and at $\alpha = 1/2$ we have 
\begin{equation}\label{eq:rintc}
    r_\text{int} \approx 0.093~\text{pc} 
    \Big(\frac{M}{M_\odot}\Big)^{0.5} \Big(\frac{L}{100~\text{pc}}\Big)^{0.5} \Big(\frac{V_L}{10~\text{km s}^{-1} }\Big)^{-1}.
\end{equation}
Obviously, 
at $r<r_\text{int}$ most pairs of stars can be gravitationally bound and form wide binaries.
With the increase of $v_\text{tur}$ with $r$,
at $r>r_\text{int}$, only a small fraction of pairs can form binaries.

%\blue{(comment on the physical meaning of the intersection?)} 

%We see that 
%with smaller $M$ and $\alpha$, 
%$v_\text{tur}$, where the PDF of $v_{12,0}$ peaks, becomes larger than $v_\text{bon}$ at a smaller $r_{12,0}$ (Fig. \ref{fig: m2a3})
%than the case with larger $M$ and $\alpha$ (Fig. \ref{fig: m10a2}). \blue{(meaning that wide binaries can form with larger binary separations?)}

\subsection{Eccentricity distribution}

In this section we calculate the eccentricity distribution that results from random pairing of stars with relative velocities drawn from the distribution in Eq. \eqref{eq: 3dveldis}.

%We consider randomly oriented binaries.
{For a bound two-body system,}
the specific energy and angular momentum are defined as 
\begin{equation}
 E = \frac{1}{2} v^2 - \frac{GM}{r}, ~~\,\,\,\,\,\,\,
    J = rv \sin \theta,
\end{equation}
where $\theta$ is the angle between 
{the separation vector}
${\bm r}$ and {the relative velocity} ${\bm v}$,
and $E$ and $J$ are related to the
eccentricity $e$ by 
\begin{equation}\label{eq: relane}
 J^2 = (1-e^2) \frac{G^2 M^2}{{(- 2 {E})}}.
\end{equation}
{Eliminating $E$ and $J$ from the above expressions leads to} 
\begin{equation}\label{eq:sinthe}
    \sin\theta = \frac{\sqrt{1-e^2}}{2 u \sqrt{1-u^2}} ,
    {\rm ~with~} u\equiv v/v_\text{bon}(r).
\end{equation}
Note that $u \in (0,1)$ for bound binaries. 
The condition $\sin\theta \in (0,1)$ further constrains the range of $u$ as 
\begin{equation}\label{eq:intbous}
      \sqrt{\frac{1-e}{2}} <u < \sqrt{\frac{1+e}{2}}
\end{equation}
at a given $e$.

{Now we {imagine that our random pairing process forms an ensemble of 
randomly oriented binaries of fixed $M$ at a given initial separation $r$. The resulting number density of these binaries}
%within $du de$ is 
in $(u, e)$ space satisfies}
(Eq. \eqref{eq:sinthe}),
\begin{equation}
\begin{aligned}
dN &\propto f(v)  dv d \cos\theta \\
   & \propto  \frac{ f(u)  du }{u\sqrt{1-u^2}\sqrt{e^2 - (2u^2-1)^2}}e de ,
\end{aligned}
\end{equation}
{where $f(u)$ is the distribution function of $u$.}
%$f(u) = v_\text{bon} f(v)$.
%are the 3D distribution functions of velocity and velocity normalized by $v_\text{bon}$, respectively.
Therefore, we find 
the distribution function of $e$ at a given $r$ as 
\begin{equation}\label{eq: orife}
\red{    p(e) =C \, e \bigintssss_{\sqrt{(1-e)/2}}^{\sqrt{(1+e)/2}} \frac{ f(u)    du }{u\sqrt{1-u^2}\sqrt{e^2 - (2u^2-1)^2}},
}
\end{equation}
where $C$ is a normalization constant, and the integral bounds are given by Eq. \eqref{eq:intbous}.
%{As a special case,} when 
%Case (i): For $f(u)$ concentrated at $u = \sqrt{1/2}$, i.e., 
%$f(u) = \delta (u-\sqrt{1/2})$, Eq. \eqref{eq: orife} becomes
%\begin{equation}
 %   p(e) = 1.
%\end{equation}
%corresponding to a constant $f(e)$.
%\red{$u = \sqrt{1/2}$ corresponds to the circular orbit speed. For an ensemble of isotropically oriented binaries considered here, 
%we naturally expect a uniform $p(e)$.}
%A uniform $p(e)$ is found for binaries with $r\sim 10^2$ AU
%which may form from disk fragmentation
%\citep{Hwang21t}.

\begin{figure*}[ht]
\centering
\subfigure[]{
\includegraphics[width=8.7cm]{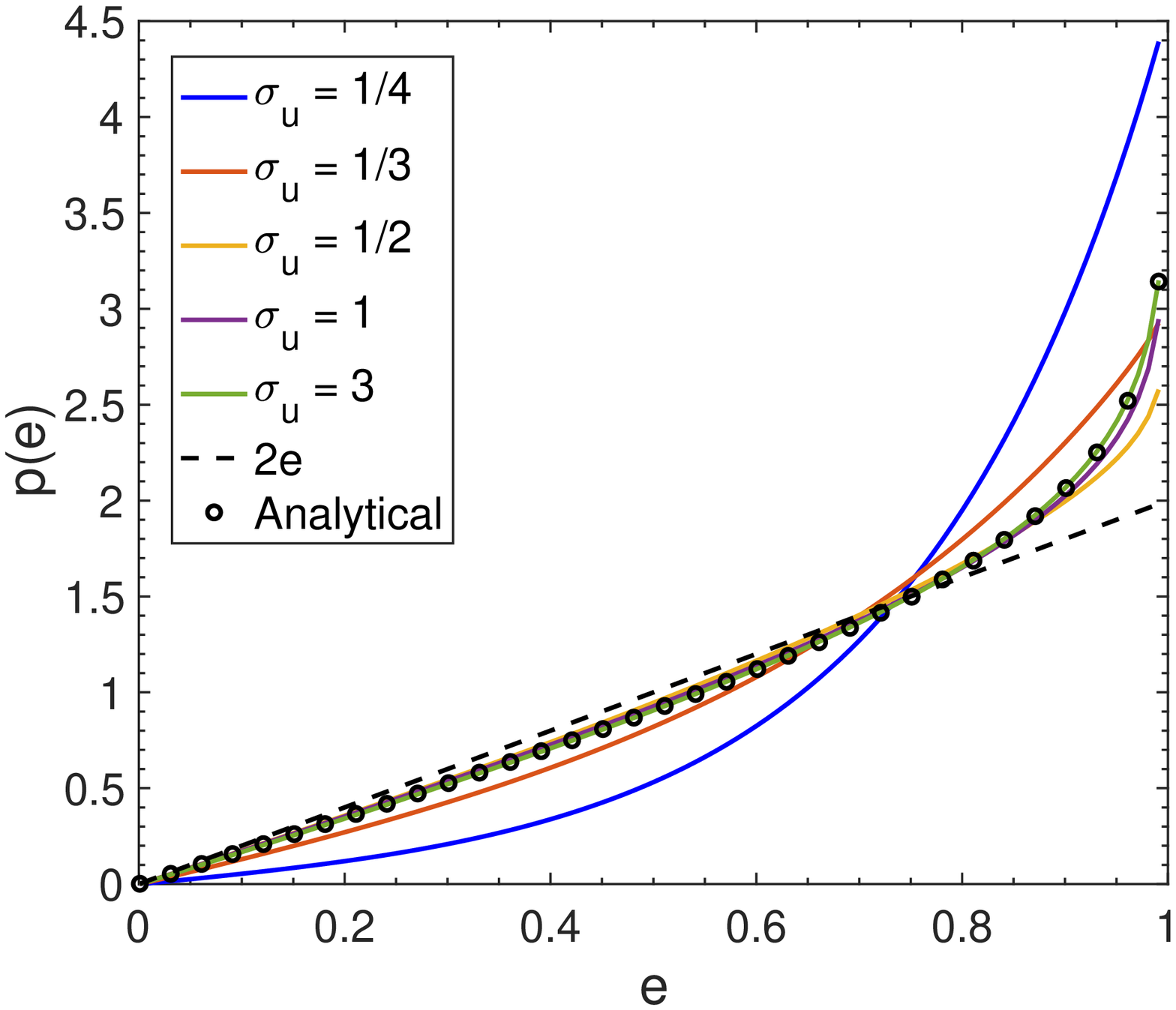}\label{fig:fumusig}}
\subfigure[]{
\includegraphics[width=8.7cm]{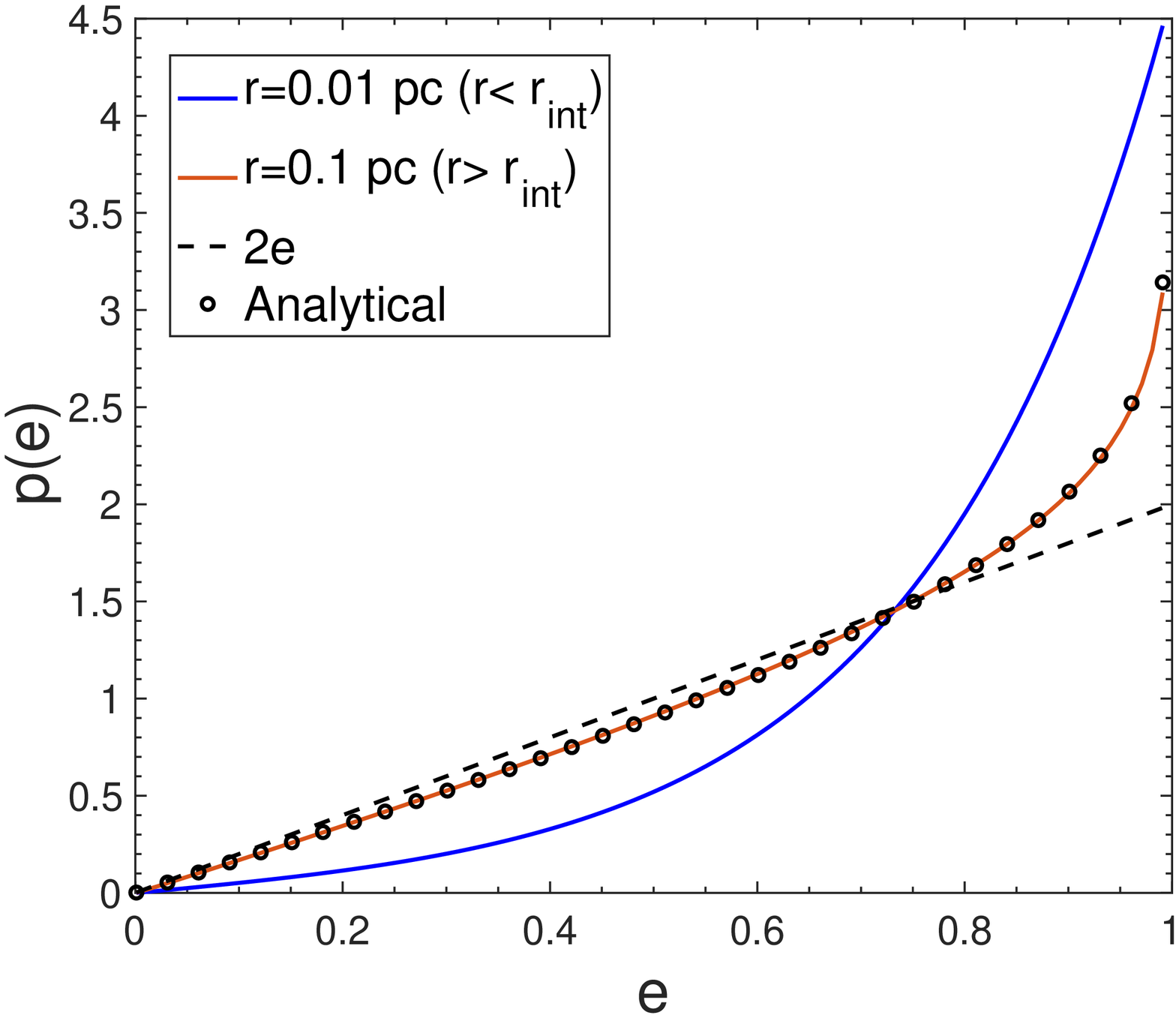}\label{fig:feana}}
\caption{(a) $p(e)$ calculated using Eq. \eqref{eq: orife}, where $f(u)$ takes the form in Eq. \eqref{eq: 3dveldis} with 
different values of $\sigma_u$ 
%(see Eq. \eqref{eq:fudis}) 
as labeled. 
The dashed line represents the thermal distribution ($p(e)=2e$). 
The circles correspond to the analytical estimate for $\sigma_u \gtrsim 1$ given by Eq. \eqref{eq:fecontu}. 
(b) Same as (a) but 
with $\sigma_u=\sigma_v/v_\text{bon}(r)$ given by $\sigma_v=0.7 v_\text{tur}$ and Eq. \eqref{eq: genturind}, using $M=2 M_\odot$, 
$L = 100$ pc, $V_L = 10$ km s$^{-1}$, and $\alpha = 1/3$.
Two different values of $r$'s (as labeled) are considered, with $r < r_\text{int}$ and $r>r_\text{int}$ (see Eq. \eqref{eq:rint}) respectively. }  %\cmtch{Panel (a) is not very informative, and in fact doesn't even look like it is correctly normalized --- perhaps using different colors would help.  Also, we should give the numbers used for the velocity DF --- `following the Gaussian' is too vague.}}
\end{figure*}

{Obviously, $p(e)$ depends on the shape of $f(u)$ within the range of integration. 
%$f(u)$ peaks at $u<1$ at $r< r_\text{int}$ 
%and $u>1$ at $r> r_\text{int}$
%(see Fig. \ref{fig: vr12}).
As an illustration, Fig. \ref{fig:fumusig} shows $p(e)$ (Eq. \eqref{eq: orife}) with 
\red{$f(u)$ taking the form 
of Eq. \eqref{eq: 3dveldis}
%\begin{equation}\label{eq:fudis}
%f(u) = 
%\frac{1}{\sigma_u\sqrt{2\pi}}\exp{\bigg(-\frac{1}{2}\Big(\frac{u-\mu_u}{\sigma_u}\Big)^2\bigg)},
%\end{equation}
for various values of  $\sigma_u = \sigma_v/v_\text{bon}$.} 
When $\sigma_u$ is much less than unity,
$p(e)$ has an excess (compared to the thermal distribution $p(e)=2e$) at large $e$'s and a deficiency at small $e$'s (see blue line in Fig. \ref{fig:fumusig}). The deficiency at small $e$'s is caused by the small $f(u)$ near $u=\sqrt{1/2}$
(corresponding to the circular orbit speed).}

%Depending on the shape of $f(u)$ within the range of integration, we can obtain the asymptotic expressions of $f(e)$. 

%Case (ii): For $f(u)$ concentrated at small $u$ with $u < \sqrt{1/2}$,
%\begin{equation}
%\begin{aligned}
%    f(e)&\approx C \bigintssss_{\sqrt{(1-e)/2}}^{u_\text{max}} \frac{ u du}{\sqrt{e^2 - (2u^2-1)^2}}e \\
%  & = C \bigg[\sin^{-1} \bigg(\frac{2u_\text{max}^2-1}{e}\bigg)+\frac{\pi}{2}\bigg]e
%\end{aligned}
%\end{equation}
%for the range $e\in(e_\text{min},1)$,
%where $u_\text{max}$ is the maximum $u$, and the corresponding minimum $e$ is $e_\text{min} = 1 - 2 u_\text{max}^2$.

%{\bf For $f(u)$ that spreads over $0<u\lesssim \sqrt{1/2}$,}
%{When $\mu_u + \sigma_u \sim \sqrt{1/2}$,} 
%Eq. \eqref{eq: orife} approximately becomes 
%\begin{equation}\label{eq:fesmu}
%    p(e) \approx C \bigintssss_{\sqrt{(1-e)/2}}^{\sqrt{1/2}} \frac{  u du }{\sqrt{e^2 - (2u^2-1)^2}}e  
%   % =C\int_{-1}^0 \frac{dy}{\sqrt{1-y^2}} e 
%    = 2e,
%\end{equation}
%which is the thermal eccentricity distribution 
%\citep{Jeans19}.
%{In Fig. \ref{fig:fumusig}, when $\mu_u + \sigma_u \sim \sqrt{1/2}$, we find that the resulting $p(e)$ is close to the thermal distribution 
%(see red lines in Fig. \ref{fig:fumusig}).} \blue{(HCH: which panel? Maybe we just add a case where $\mu_u=\sigma_u=0.5 \sqrt{1/2}$?)}

{When $\sigma_u$ is larger than unity,} 
we \red{approximately have $f(u) \propto u^2$} over $0<u<1$. %\blue{(HCH: not true if we adopt speed distribution)}.
{In this case, Eq. \eqref{eq: orife} simplifies to 
\begin{equation}\label{eq:fecontu}
\begin{aligned}
 %f(e) = C\sqrt{e}\bigintssss_{-1}^1 \frac{dy}{\sqrt{\frac{1}{e}-y} \sqrt{1-y^2}},
 p(e)&\approx 
 %C \bigintssss_{(1-e)/2}^{(1+e)/2} \frac{dy}{\sqrt{1-y} \sqrt{\frac{1+e}{2}-y}\sqrt{y-\frac{1-e}{2}}} e \\
% &=  \frac{Ce}{\sqrt{1+e}} F \Bigg(\frac{\pi}{2}, \sqrt{\frac{2e}{1+e}}\Bigg)
%& = 
\frac{1.06\, e}{\sqrt{1+e}} K\Bigg(\sqrt{\frac{2e}{1+e}}\Bigg),
\end{aligned}
\end{equation}
%where $y = (2u^2-1)/e$.
where 
%$y = u^2$, 
%\begin{equation}
%   C = \frac{1}{\int_0^1 \frac{e}{\sqrt{1+e}} K\Big(\sqrt{\frac{2e}{1+e}}\Big)de},
%\end{equation}
%and 
{$K$ is the complete} elliptic integral of the first kind
\citep{GraR94}.} 
As shown in Fig. \ref{fig:fumusig}, 
the above expression corresponds to a superthermal $p(e)$ and 
agrees well with the cases of large $\sigma_u$ (purple and green lines). %\blue{(HCH: blue lines?)}

In Fig. \ref{fig:feana}, we present $p(e)$ 
(Eq. \eqref{eq: orife}) with 
$f(u)$ taking the form 
of Eq. \eqref{eq: 3dveldis}
for 
$M=2 M_\odot$, 
$L = 100$ pc, $V_L = 10$ km s$^{-1}$, and $\alpha = 1/3$.
%To illustrate the effect of velocity distribution on $f(e)$, we also present $f(e)$ with a Maxwellian velocity distribution as a comparison, for which we take $v_\text{tur}$ as the most probable speed.
At $r =0.01$ pc with $r < r_\text{int}$ (see Eqs. \eqref{eq:rintk} and \eqref{eq:rintc}, 
Fig. \ref{fig: vr12}), 
it falls in the regime where $\sigma_u$ is much less than unity. Therefore, we see a deficiency at small $e$'s and
a significant excess at large $e$'s.
%The result for a Maxwellian velocity distribution deviates from the thermal eccentricity distribution due to the excess of $f(u)$ near $u\sim 0$ and the deficit of $f(u)$ near $u\sim \sqrt{1/2}$.
%With small $f(u)$ near $u=0$ and $u=1$, $f(e)$ for both Gaussian and Maxwellian
%velocity distributions deviates from that for a constant $f(u)$.
{At $r =0.1$ pc with $r > r_\text{int}$
%The Gaussian velocity distributions at $v\ll \mu_v$ is approximately constant. 
and $\sigma_u$ larger than unity,} 
the corresponding $p(e)$ is well described by Eq. \eqref{eq:fecontu} and is superthermal.
%The transition from thermal to superthermal $p(e)$ with the increase of $r$ 
%is consistent with the observational finding in e.g., \citet{Toko20,Hwang21t}.
\red{We see that irrespective of the value of $\sigma_u$, a superthermal $p(e)$ is generally expected. }

\begin{figure}[ht]
\centering
%\subfigure[]{
%   \includegraphics[width=8.7cm]{ecendis.eps}\label{fig: ece}}
%\subfigure[]{
%   \includegraphics[width=8.7cm]{fecomred.eps}\label{fig: fecom}}
\includegraphics[width=9cm]{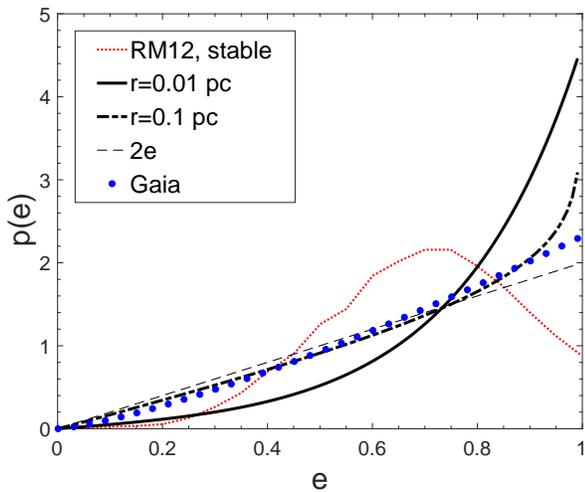}
\caption{
%(a) $f(e)$ for binary stars with initial single velocity $v_{12,0}= \beta v_c$ and $\beta =0.7$ (dash-dotted line, Eq. \eqref{eq: sigvfe}) and with initial
%turbulent velocities 
%(solid line, Eq. \eqref{eq: supthe}), in comparison with the thermal eccentricity distribution (dashed line). 
Comparison between $p(e)$ derived in this work (solid and dash-dotted lines, same as in Fig. \ref{fig:feana}) and that taken from \citet{Reip12} (RM12). 
Red dotted line represents the normalized $p(e)$ taken from
\citet{Reip12}
for the outer binaries %\cmtch{What does `the outer pairs' mean.?  And the superhtermal DF we are trying to reproduce is not contaminated by triples anyway, is it?} 
in stable bound triples at $1$ Myr, with 
the distribution of semimajor axes of outer binaries peaked around $10^3$ AU.
Dashed line is the thermal distribution. 
Blue circles show a power-law fit $p(e) \propto e^{1.32}$ to  
{\it Gaia} wide binaries with separations $10^{3.5}-10^4$ AU
in the Galactic field
\citep{Hwang21t}.   }
\label{fig:com}
\end{figure}

In Fig. \ref{fig:com},
we compare our $p(e)$ of wide binaries taken from Fig. \ref{fig:feana}
and that 
of the wide binaries formed from dynamical unfolding of triple systems at $1$ Myr taken from 
\citet{Reip12}.
%\footnote{Unlike our mechanism, as many triple systems are not unfolded fully at $1$ Myr
%\citep{Reip12}, 
%most binaries in triples
%do not appear as wide binaries in a young star-forming region.}
In the latter case, $p(e)$ for stable bound triples declines at large $e$'s.
%while $p(e)$ for unstable bound triples is superthermal, but such systems can be easily disrupted and short-lived
%\citep{Reip12}.
A superthermal $p(e)$ of the wide binaries {at binary separations $>10^3$\,AU} in the Solar neighbourhood is indicated by {\it Gaia} observations
\citep{Hwang21t} (see Fig. \ref{fig:com}).
Compared with the observations, we see a more significant 
excess at large $e$'s that we derive for young wide binaries. This excess is likely to be 
reduced by other physical processes
that may occur after the binary formation, e.g. dynamical interaction/scatterings between
binaries and passing stars and MCs. 
%in star-forming clusters
%\blue{(For these comparisons, we need to specify the semi-major axis range we are comparing with. For wide binaries in Gaia, I guess the range is $10^3$-$10^4$\,AU, or something like that. But I'm not sure the range of semi-major axes from Reipurth?)}. 
%Our result reconciles better with the observations, 
{The superthermal $p(e)$ we find may therefore be important for understanding the observations, especially since those observations likely reflect the formation process of wide binaries rather than their subsequent dynamical interactions \citep{Hamil22}.}
%Because the ultra wide binaries with large $e$ are most vulnerable to 
%external perturbations in the Galactic environment 
%\blue{(this is a more complicated process. I may just say that, "the further interaction with passing stars and Galactic tide makes the eccentricity distribution closer to the thermal eccentricity distribution")}, 
%the difference at the high-$e$ end between our result of newly formed wide binaries 
%and the observed distribution in the field 
%is well expected. 
%Future measurements on $f(e)$ of young binaries ($\lesssim$ 1 Myr old) 
%can be used to distinguish between our model and \citet{Reip12}
%model
%based on the different theoretical expectations on
%$f(e)$.  

\section{Discussions}

Observationally, wide binaries 
at separations $>10^3$\,AU 
in the Solar neighbourhood
have a superthermal eccentricity distribution \citep{Hwang21t}. Since the effect of Galactic tides cannot produce the superthermal eccentricity distribution on its own 
{and diffusive scattering with passing stars and MCs would tend to make the distribution more thermal}
\citep{Hamil22}, the observed eccentricity distribution 
{is likely a relic of the wide binary formation mechanism.}

Simulations show that wide binaries formed from cluster dissolution would have a thermal eccentricity distribution \citep{Kouw10} and therefore cannot explain the observation. While wide tertiaries formed from dynamical unfolding of compact triples are predicted to be highly eccentric \citep{Reip12}, \cite{Hwang23} finds that the eccentricities of wide tertiaries in triples are similar to the wide binaries at the same separations, suggesting that the dynamical unfolding scenario plays a minor role. The remaining formation channels that form eccentric wide binaries are turbulent fragmentation \citep{Bat98}
and random pairing during star formation phase \citep{Toko17}.
%\red{and tidal and collisional effects during the early evolutionary phases of ultra-faint dwarf galaxies
%\citep{Liver23}.} 
Simulations of turbulent fragmentation suggest that wide binaries at $>10^3$\,AU are eccentric ($e>0.6$) \citep{Bate2014}, although the number of binaries in simulations is too low to have well-characterized eccentricity distributions. 

In this paper, we have focused  on the {eccentricity distribution that results} from the random pairing scenario 
{under the \red{consideration that the initial relative velocities of pairs of stars are likely drawn from a characteristic turbulent velocity distribution of the star-forming gas}.}
%For binaries with semi-major axes $a<r_{\rm bon}$, they would be highly eccentric, and at $a\sim r_{\rm bon}$, the eccentricity distribution would be slightly superthermal. 
Given the typical {turbulence conditions} 
%densities and velocity dispersion 
in star-forming regions, wide binaries formed from random pairing are expected to be highly eccentric, which may be (partly) responsible for the observed superthermal eccentricity. 

We note that while the random pairing scenario may explain the high-eccentricity wide binaries, 
{it is not clear whether it} 
%it is unlikely to 
dominates the wide binary formation at $>10^3$\,AU. %First, {\bf under the simplified assumption of uniform $n$,}
%random pairing scenario leads to the increasing binary fraction with increasing separations 
%%up to $r_{\rm int}$ 
%{in a star-forming region}
%(Section \ref{ssec:binf}), but observationally the binary fraction
%{in the field}
%peaks around $50$\,AU and decreases towards larger binary separations \citep{Duquennoy1991,Raghavan2010,El-Badry2018}. Second, 
Wide binaries formed from random pairing are weakly bound and their further gravitational interactions with other stars 
%in the clusters
may disrupt the binaries. However, even if random pairing only contributes a few per cent of wide binaries at $>10^3$\,AU, they may still significantly change the eccentricity distribution to superthermal (e.g. \citealt{Hwang2022twin}).

%For a turbulent velocity field, 
%while the velocity increments at large separations are Gaussian distributed,
%deviations from the Gaussian distribution with large-amplitude tails are seen for the velocity longitudinal increments at small separations
%\citep{Fris95,Falc14}.
%\red{Irrespective of the exact shape of $f(u)$,
%a superthermal $p(e)$ is generally expected. }

\section{Summary}

Turbulence plays a fundamental role in star formation.
The turbulent motions in molecular gas are inherited by stars at their birth. 
Different from previous studies relying on the long-term dynamical evolution of stars and star clusters, 
we suggest that the formation of wide binaries with initial separations in the range
$0.01~\text{pc} \lesssim r\lesssim 1$ pc (i.e., $2\times10^3~\text{AU} \lesssim r \lesssim 2\times10^5$~AU)
is a natural consequence of star formation in the turbulent interstellar medium. 
With the velocity differences and separations of newly formed stars statistically following the turbulent velocity scaling, 
a pair of stars with a sufficiently small velocity difference at a given separation is 
gravitationally bound and can form a wide binary.

%The energy of the binary system depends on the scale-dependent turbulent kinetic energy, 
%and a larger $a$ is expected at a larger initial separation with a larger turbulent energy. 
%As a result, wide binaries formed in a turbulent MC have a wide range of $a$. %and $T$, 
%as indicated by observations 
%(e.g., \citealt{Krou95,Toko17}). %\blue{(HCH: this comparison with observations may be dangerous because we are only probing the widest binaries from a specific channel (turbulence), while the wide range of a and T in binaries is likely due to different formation channels.)}

\red{For a turbulent speed distribution of pairs of stars, we find that the resulting eccentricity distribution of the bound pairs is generally superthermal.  This is true regardless of whether the average turbulent velocity $v_\text{tur}$ is smaller or larger than the escape velocity $v_\text{bon}$ of a gravitationally-bound binary}.
%The separation $r_\text{int}$ corresponds to the intersection between the averaged turbulent velocity $v_\text{tur}$ and the binding velocity of a gravitationally-bound binary $v_\text{bon}$. It depends on the turbulent scaling and $M$.
%The eccentricity distribution $f(e)$ of wide binaries at a given separation depends on the turbulent velocity distribution. When a Gaussian velocity distribution is considered, with the separation smaller than $r_\text{int}$, $f(e)$ is close to the thermal distribution with a slight excess at large eccentricities. 
%With the separation larger than $r_\text{int}$, $f(e)$ is superthermal. 
%The transition from thermal to superthermal $f(e)$ with increasing separations of wide binaries is also indicated by {\it Gaia} observations 
%(e.g., \citealt{Toko20,Hwang21t}).
The superthermal $p(e)$ of wide binaries {under this formation channel may} be important for explaining the observed superthermal $p(e)$ in the Solar neighbourhood.
%which is unlikely to develop from dynamical perturbations induced by, e.g., passing stars and MCs,  Galactic tides
%\citep{Hamil22}.
{Comparisons with future measurements on $p(e)$ in star-forming regions will provide testing of our theory.

\acknowledgments
S.X. acknowledges the support for 
this work provided by NASA through the NASA Hubble Fellowship grant \# HST-HF2-51473.001-A awarded by the Space Telescope Science Institute, which is operated by the Association of Universities for Research in Astronomy, Incorporated, under NASA contract NAS5-26555. HCH acknowledges the support from the Infosys Membership at the Institute for Advanced Study.
{This work was supported by a grant from the Simons Foundation (816048, CH).}

\software{MATLAB \citep{MATLAB:2021}}

\bibliographystyle{aasjournal}
\bibliography{xu}

\end{document}